\documentstyle[12pt,epsf]{article}

\begin{document}

\hfill
\vtop{\hbox{\bf MADPH-00-1170}
      \hbox{\bf Fermilab-PUB 00/081-T} \hbox{\bf AMES-HET 00-02}
      \hbox{April 2000}}

\vspace{.5in}

\begin{center}
{\large\bf Determination of the Pattern of Neutrino Masses\\[2mm] at a
 Neutrino Factory }\\[6mm]

V. Barger$^1$, S. Geer$^2$, R. Raja$^2$, and K. Whisnant$^3$\\[3mm]
\it
$^1$Department of Physics, University of Wisconsin, Madison, WI 53706,
USA\\ $^2$Fermi National Accelerator Laboratory, P.O. Box 500,
Batavia, IL 60510, USA\\ $^3$Department of Physics and Astronomy, Iowa
State University, Ames, IA 50011, USA

\end{center}

\vspace{.5in}

\begin{abstract}
We study  the precision with which the sign of  $\delta m_{32}^2$ can
be determined at a neutrino factory, as a function of stored muon energies
and baselines. This is done by  simultaneously fitting the channels
$\nu_\mu \rightarrow \nu_\mu$\,, $\bar\nu_e \rightarrow \bar\nu_\mu $
from $\mu^-$ decays and the channels $\bar\nu_\mu \rightarrow
\bar\nu_\mu$\,, $\nu_e \rightarrow \nu_\mu $ from $\mu^+$ decays. 
For  a 20~GeV muon storage ring we investigate as a function 
of the baseline length and the magnitude of $\delta m_{32}^2$, 
the minimum 
value of the parameter $\sin^22\theta_{13}$ for which 
the sign of $\delta m_{32}^2$ can be determined to at least 3 standard 
deviations. We find that for baselines longer than $\sim3000$~km,
the sign of $\delta m^2_{32}$ can be determined for $\sin^22\theta_{13}$
down to the $10^{-3}$ level with $10^{20}$ decays of 20~GeV muons.

\end{abstract}

\thispagestyle{empty}
\newpage

\section{Introduction}

The conceptual development of very intense muon
sources~\cite{status_report} has led to a proposal~\cite{geer} to use
this new accelerator technology to build a Neutrino Factory in which
an intense low energy muon beam is rapidly accelerated and injected
into a storage ring with long straight sections. The muons decaying in
these straight sections produce intense beams of highly collimated
neutrinos. The technical and physics possibilities for building and
using neutrino factories have been explored by many
groups~\cite{fnal_accel_study,geer,many}. The interest in neutrino factories
is primarily driven by recent measurements from the SuperKamiokande
(SuperK) collaboration~\cite{superk}, which indicate that muon neutrinos
produced in
atmospheric interactions of cosmic rays oscillate into other neutrino
flavors, a result that is consistent with measurements made in other
experiments~\cite{kam}. The presence of a well-defined electron
neutrino flux in the muon storage ring neutrino beams permits one to
explore the effect of matter on the propagation of electron
neutrinos. The analysis of the oscillation data leads to several scenarios
of neutrino masses and mixing.  In this paper, we investigate the
effect of passage through matter on neutrino oscillations. We specifically
consider the Large
Angle MSW scenario (LMA)~\cite{solarlam}, which, as defined in the
recent Fermilab six month physics study~\cite{fnal_physics_study}, has
\begin{eqnarray}
& |\delta m_{32}^2| = 3.5\times10^{-3}\,{\rm eV}^2, \qquad |\delta
m_{21}^2| = 5\times10^{-5} {\rm eV}^2, & \nonumber\\ &
\sin^22\theta_{23}=1.0, \qquad
\sin^22\theta_{12}=0.8,  \qquad
\sin^22\theta_{13}=0.04 &   \label{eq:lam}
\end{eqnarray}
and the CP violating phase $\delta = 0$.  However, to the extent that the
contributions from the subleading solar $\delta m_{21}^2$ scale are small, our
results apply approximately to other solar scenarios. The transition
probabilities
in the leading oscillation approximation for propagation through
matter of constant density are~\cite{BGRW,wolf-etal,pantaleone,gomez,rubbia}
\begin{eqnarray}
P(\nu_e\to \nu_\mu) &=& s_{23}^2 \sin^2 2\theta_{13}^m
\sin^2\Delta_{32}^m \,,
\nonumber\\
P(\nu_e\to \nu_\tau) &=& c_{23}^2 \sin^2 2\theta_{13}^m
\sin^2\Delta_{32}^m \,,
\label{eq:probs}\\
P(\nu_\mu\to \nu_\tau) &=& \sin^2 2\theta_{23} \left[
(\sin\theta_{13}^m)^2 \sin^2\Delta_{21}^m +(\cos\theta_{13}^m)^2
\sin^2\Delta_{31}^m -(\sin\theta_{13}^m\cos\theta_{13}^m)^2
\sin^2\Delta_{32}^m \right]
\,. \nonumber
\end{eqnarray}
 The oscillation arguments are given by
\begin{equation}
\Delta_{32}^m = \Delta_0 S \,,\qquad
\Delta_{31}^m = \Delta_0 {1\over2} \left[ 1+{A\over\delta m^2_{32}}+S \right]
\,, \qquad
\Delta_{21}^m = \Delta_0 {1\over2} \left[ 1+{A\over\delta m^2_{32}}-S \right]
\,, \label{eq:arg}
\end{equation}
where $S$ is given by
\begin{equation}
S \equiv \sqrt{ \left( {A\over\delta m^2_{32}}-\cos2\theta_{13}
\right)^2 + \sin^2 2\theta_{13}} \,,
\label{eq:S}
\end{equation}
and
\begin{equation}
\Delta_0 = {\delta m^2_{32} L\over 4E} = 1.267 {\delta m^2_{32}
{\rm\,(eV^2)} \; L {\rm\ (km)} \over E_\nu {\rm\ (GeV)}}
\,.
\label{eq:arg0}
\end{equation}
\begin{equation}
\sin^2 2\theta_{13}^m = {\sin^2 2\theta_{13}\over
\left({A\over\delta m^2_{32}} - \cos 2\theta_{13} \right)^2
+ \sin^2 2\theta_{13}} \,. \label{eq:sin}
\end{equation}
The amplitude $A$ for $\nu_e e$ forward scattering in matter is given
by
\begin{equation}
A = 2\sqrt2 G_F N_e E_\nu = 1.52 \times 10^{-4}{\rm\,eV^2} Y_e
\rho({\rm\,g/cm^3}) E({\rm\,GeV}) \,.
\label{eq:A}
\end{equation}
Here $Y_e$ is the electron fraction and $\rho(x)$ is the matter
density. For neutrino trajectories that pass through the earth's
crust, the average density is typically of order 3~gm/cm$^3$ and $Y_e
\simeq 0.5$.  The oscillation probability $P(\nu_e\to \nu_\mu)$ is
directly proportional to $\sin^2 2\theta_{13}^m$, which is
approximately proportional to $\sin^2 2\theta_{13}$. There is a
resonant enhancement for
\begin{equation}
  \cos2\theta_{13} = \frac{A}{\delta m^2_{32}} \,.
\end{equation}
For electron neutrinos, $A$ is positive and the resonance enhancement
occurs for positive values of $\delta m^2_{32}$ for $\cos2\theta_{13}>0$.
 The reverse is true
for electron anti-neutrinos and the enhancement occurs for negative
values of $\delta m^2_{32}$.  Thus for a neutrino factory operating
with positive stored muons (producing a $\nu_e$ beam) one expects an
enhanced production of opposite sign ($\mu^-$) charged-current events
as a result of the oscillation $\nu_e\to \nu_\mu$ if $\delta m^2_{32}$
is positive and vice versa for stored negative
beams~\cite{BGRW,gomez,rubbia,shrock,freund,bernstein-etal}.

\begin{figure}
\epsfxsize=6in\epsffile{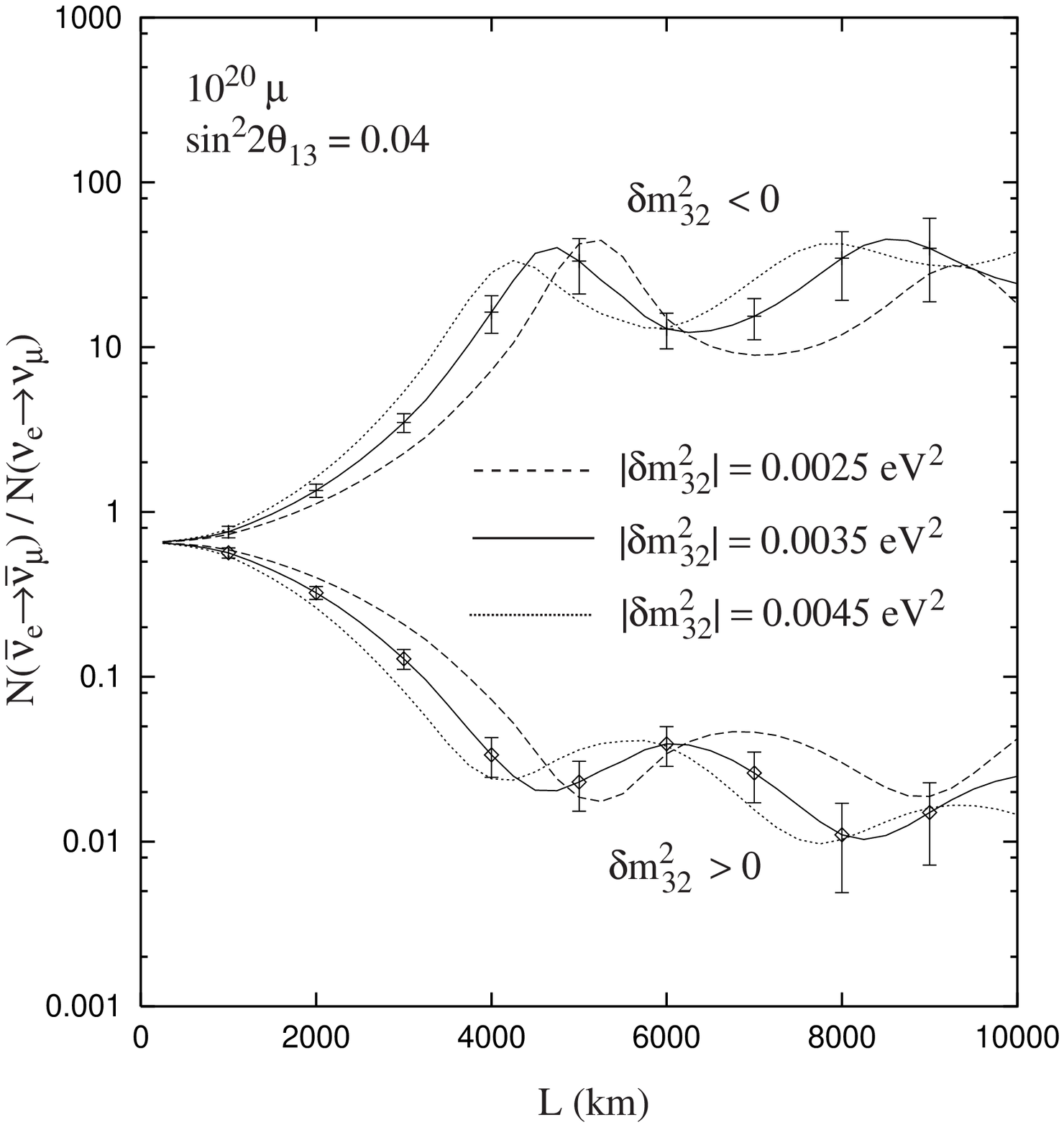}
\caption[]{
The ratio of wrong-sign muon event rates $N(\bar\nu_e \to \bar\nu_\mu) /
N(\nu_e \to \nu_\mu)$ versus baseline for a 20 GeV muon storage ring
and $\delta m^2_{32} = 0.0025$~eV$^2$ (dashed line), $0.0035$~eV$^2$
(solid line), and $0.0045$~eV$^2$ (dotted line). The other oscillation
parameters are given in Eq.~(\ref{eq:lam}). A 4~GeV minimum cut was
imposed on the detected muon energy. The error bars show the statistical
errors corresponding to 10$^{20}$ decays and a 50~kiloton detector.
\label{fig:ratios}}
\end{figure}

This enhancement is evident in Fig.~\ref{fig:ratios}, which shows the
ratio of $\bar\nu_e \to \bar\nu_\mu$ events from $\mu^-$ decays to
$\nu_e \to \nu_\mu$ events from $\mu^+$ decays for 20~GeV muons and a
50~kt detector, assuming the oscillation parameters of
Eq.~(\ref{eq:lam}). The results for two other values of $\delta
m^2_{32}$ are also presented in Fig.~\ref{fig:ratios}. This figure
shows that for larger $L$ the ratio of wrong-sign muon events
is sensitive to the sign of $\delta m^2_{32}$.

The magnitude of $\delta m^2_{32}$ is determined from the disappearance
of muon neutrinos due to the oscillation $\nu_\mu\to \nu_\tau$, since it
can be shown that for the baselines under consideration here, the matter
effects are small in this channel (see Fig.~2, curves 2 and 3 of
Ref.~\cite{BGRW}).  The oscillation probability$ P(\nu_\mu \rightarrow
\nu_\tau)$ can thus be approximated by $\cos^4\theta_{13}
\sin^22\theta_{23} \sin^2(\Delta_0)$ as though it were in vacuum for
baselines as far as 4000~km. (See the section ``Method" for a correction
to this approximation for matter effects for longer baselines.)

\section{Method}

In order to extract both the sign and magnitude of $\delta m^2_{32}$,
we simultaneously fit four channels
(i) $\nu_\mu \rightarrow \nu_\mu$
(ii) $\bar{\nu_e} \rightarrow \bar{\nu_\mu}$
(iii) $\bar{\nu_\mu} \rightarrow \bar{\nu_\mu}$
(iv) $\nu_e \rightarrow \nu_\mu$. The first two channels are
measured when N $\mu^-$ decays occur in the storage ring and the second two
channels
are observed when N $\mu^+$ decays occur~\cite{decays}.

We study the extraction of oscillation parameters in two scenarios:
\begin{itemize}
\item  A  muon storage ring with $N=10^{20}$ muon decays
 and a 50 kiloton detector.
\item  A  muon storage ring with $N= 10^{19}$ muon decays  and
a 50 kiloton detector. A 20~GeV version of this is known as an ``entry
level'' neutrino factory.
\end{itemize}

\begin{figure}
\epsfxsize=6in\epsffile{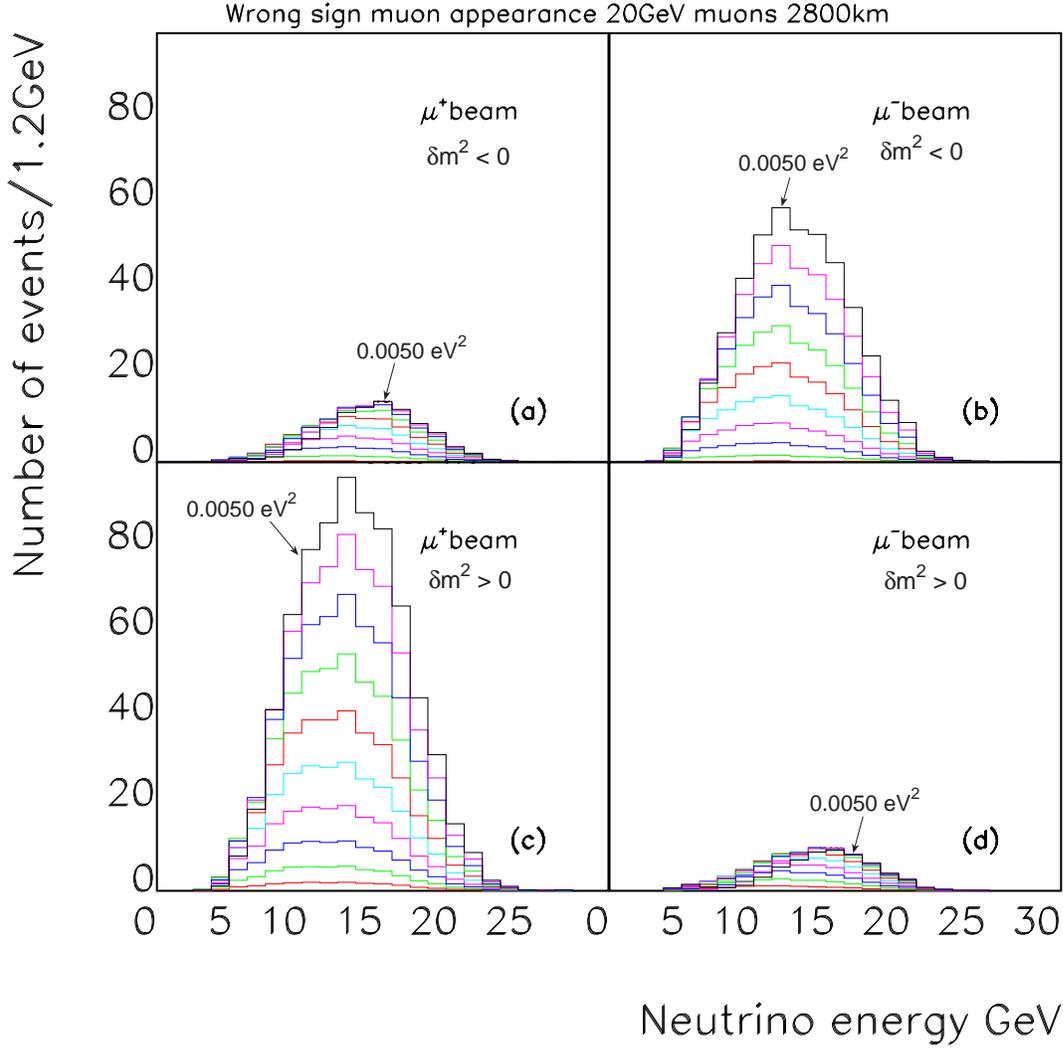}
\caption[]{
The wrong sign muon appearance rates for a 20 GeV muon storage ring at
a baseline of 2800~km with 10$^{20}$ decays and a 50 kiloton detector
for (a)~$\mu^+$ stored and negative $\delta m^2_{32}$\,, (b)~$\mu^-$ stored
and negative $\delta m^2_{32}$\,, (c)~$\mu^+$ stored and positive $\delta
m^2_{32}$\,,
(d)~$\mu^-$ stored and positive $\delta m^2_{32}$. The values of $|\delta
m^2_{32}|$ range from 0.0005 to 0.0050 eV$^2$ in steps of 0.0005~eV$^2$.  
Matter enhancements are evident in (b) and (c).
\label{fig:hists}}
\end{figure}

Our study is performed for baselines ranging from 732~km
up to 7332~km and stored muon energies ranging from 20~GeV to 50~GeV.
We calculate the neutrino event rates by propagating the neutrinos
through matter, taking into account the variations in the density
profile using the Preliminary Reference Earth Model~\cite{PREM} by
solving the evolution equations numerically.  Realistic detector
resolutions are used that are appropriate for a magnetized iron
scintillator detector~\cite{gomez,fnal_physics_study}. 
We assume a muon energy resolution
$\frac{\sigma}{E_\mu}=0.05$ and a hadronic shower resolution of
$\frac{\sigma}{E_h}=0.53/\sqrt{E_h}$ for showers of energy $E_h>3$~GeV and
0.8/$\sqrt{E_h}$ for showers of energy $E_h<3$~GeV. We accept only events
that possess muons of true energy greater than 4 GeV.  Figure
\ref{fig:hists} shows the wrong-sign muon appearance spectra with
these cuts as function of $\delta m^2_{32}$ for both $\mu^+$ and
$\mu^-$ beams for both signs of $\delta m^2_{32}$ at a baseline of
2800~km. The resonance enhancement in wrong sign muon production is
clearly seen in Fig.~\ref{fig:hists} (b) and (c). Using these
histograms and similar ones for the disappearance channels, it is
possible to predict the spectrum in any channel for any value of
$\delta m^2_{32}$ in the range of interest by polynomial interpolation
of the histograms bin by bin with the method of divided differences.
For the interpolation, the disappearance probabilities can be
treated to first order as
though they are due to vacuum oscillations (i.e. matter effects can be
neglected) and are proportional to
$\cos^4\theta_{13}\sin^22\theta_{23}$.  For the longer baselines, the
matter effects affect these disappearance rates slightly, as can be
seen from Fig.~3 of Ref.~\cite{BGRW}. The difference in the
disappearance rates for positive and negative $\delta m^2_{32}$ is
proportional to $\sin^2 2\theta_{13}$, vanishing as $\sin^2
2\theta_{13}\rightarrow 0$. For the channel $\nu_\mu \rightarrow
\nu_\mu$, the positive $\delta m^2_{32}$ solution approaches the negative
$\delta m^2_{32}$ solution as $\sin^2 2\theta_{13}\rightarrow 0$, the reverse
being the case for anti-neutrinos. We use this behavior to interpolate
the disappearance spectra as a function of $\sin^2 2\theta_{13}$. The
event rates in both the appearance and disappearance channels depend
only on the three vacuum parameters $\sin^2 2\theta_{23}$, $\sin^2
2\theta_{13}$, and $\delta m^2_{32}$.  Using this interpolation
scheme, we can generate events for any combination of these
parameters for various muon momenta and baselines. Most of the
information on the sign of $\delta m^2_{32}$ comes from the appearance
channels  and the precision on the magnitude of  $\delta m^2_{32}$
comes from the disappearance channels~\cite{BGRW,gomez,rubbia,freund}.

\subsection {Backgrounds}
The background for the wrong-sign muon appearance channels arise from
three sources: (i)~charm production, (ii)~$\pi, K$ decay events
producing wrong sign muons in neutral current interactions, and (iii)~$\pi, K$
decay events producing wrong sign muons in charged-current
(CC) events where the primary muon was considered
lost~\cite{fnal_physics_study,gomez}. A more detailed study of the
backgrounds~\cite{fnal_physics_study} has shown that significant
reductions are obtained by demanding that $P^2_t> 2\rm~GeV^2$, where
$P^2_t$ is defined as the transverse-momentum-squared of the muon with
respect to the hadronic shower direction. Imposing this cut results in
a further signal efficiency factor of $0.9$~\cite{spentz} for the CC
disappearance channels, $0.62$ for the $\mu^+$ beam appearance signal
and $0.45$ for the $\mu^-$ beam appearance signal.  This results in an
average background/$\nu_\mu \rightarrow \nu_\mu$ CC signal rate of $0.45\times
10^{-4}$ for $\mu^+$ beam appearance channel and $0.25\times 10^{-5}$
for the $\mu^-$ beam appearance channel.  The difference in these
rates between the $\mu^+$ and $\mu^-$ beams is due to the different
kinematical distributions of the neutrino and anti-neutrino CC
interactions. We fold these rates into the theoretical
prediction for each channel.  We also
assume a normalization systematic uncertainty of 1$\%$ between
the $\mu^+$ beam and $\mu^-$ beam events.

Table~\ref{dm2table} lists the wrong sign muon appearance-- 
and background--rates for an ``entry level'' machine. 
Event rates for the appearance channels are appreciable, whereas background 
rates are negligible for baselines longer than about 2800~km.

\subsection{Fitting}
Events are generated for the 4 channels $\nu_\mu \rightarrow
\nu_\mu$, $\bar{\nu_e}\rightarrow \bar{\nu_\mu}$, $\bar{\nu_\mu}
\rightarrow \bar{\nu_\mu}$, and $\nu_e\rightarrow
\nu_\mu$, and for both signs of $\delta m^2_{32}$.  The four 
simulated measured event rates are fitted 
simultaneously for the three parameters $\sin^22\theta_{13}$,
$\sin^22\theta_{23}$, and $\delta m^2_{32}$. In each fit the sign 
of $\delta m^2_{32}$ is constrained~\cite{raja}. The difference in negative
log-likelihood, $\Delta L$,
between the fits in which sign of $\delta m^2_{32}$ has been constrained
to the correct and incorrect sign is evaluated for a
given value of $\sin^22\theta_{13}$. This is then 
repeated for a range of $\sin^22\theta_{13}$. It is empirically found
that the average $\Delta L$ varies linearly with $\sin^22\theta_{13}$
and this is used to estimate that value of $\sin^22\theta_{13}$ at
which $\Delta L$=4.5, i.e a Gaussian 3$\sigma$ ability to
differentiate the sign of $\delta m^2_{32}$. This value of
$\sin^22\theta_{13}$ we define as the 3$\sigma$ reach in 
$\sin^22\theta_{13}$ space. We only fit for the number of events and not the
shape of the spectra, since at the $3\sigma$ point statistics are such
that shape information contributes little.

\begin{table}[h]
\caption[]{Wrong-sign muon rates for a 50~kt detector
(with a muon threshold of 4~GeV) a distance $L$ downstream of a muon
factory (energy $E_\mu$) providing $10^{19}$ muon decays. Rates are
shown for LMA scenario of Eq.~(\ref{eq:lam}) with both signs of
$\delta m^2_{32}$ considered separately. The background rates listed
are for each sign of $\delta m^2_{32}$ and do not depend on the sign
of $\delta m^2_{32}$.}
\bigskip
\def\tabcolsep{.75em}
\centering\leavevmode
\begin{tabular}{cl|lll|lll}
\hline\hline
$E_\mu$&$L$&\multicolumn{3}{c}{$\mu^+$
stored}&\multicolumn{3}{c}{$\mu^-$ stored} \\ GeV & km & $\delta
m^2_{32} > 0$ &  $\delta m^2_{32} < 0$ & Backg & $\delta
m^2_{32} > 0$ & $\delta m^2_{32} < 0$ & Backg \\
\hline
   20 &    732 &  32.5 &  22.7 &   9.6 &  14.3 &  11.6 &   0.9\\
   20 &   2800 &  28.7 &   5.7 &   0.3 &   3.2 &  11.7 &   0.0\\
   20 &   7332 &  20.4 &   0.6 &   0.0 &   0.2 &   8.5 &   0.0\\
\hline
   30 &    732 &  54.5 &  38.5 &  19.1 &  23.0 &  18.3 &   1.9\\
   30 &   2800 &  49.2 &  13.1 &   0.8 &   7.4 &  18.4 &   0.1\\
   30 &   7332 &  26.2 &   1.7 &   0.0 &   0.7 &  11.9 &   0.0\\
\hline
   40 &    732 &  56.7 &  40.2 &  17.7 &  23.6 &  18.7 &   1.7\\
   40 &   2800 &  51.3 &  14.7 &   0.8 &   8.2 &  18.7 &   0.1\\
   40 &   7332 &  26.8 &   1.9 &   0.0 &   0.8 &  11.3 &   0.0\\
\hline
   50 &    732 &  53.3 &  37.9 &  15.7 &  21.9 &  17.3 &   1.5\\
   50 &   2800 &  47.7 &  13.9 &   0.7 &   7.9 &  17.4 &   0.1\\
   50 &   7332 &  25.8 &   1.9 &   0.0 &   0.8 &  10.8 &   0.0\\
\hline\hline
\end{tabular}
\label{dm2table}
\end{table}

\section{Results}

\begin{figure}
\epsfxsize=6in\epsffile{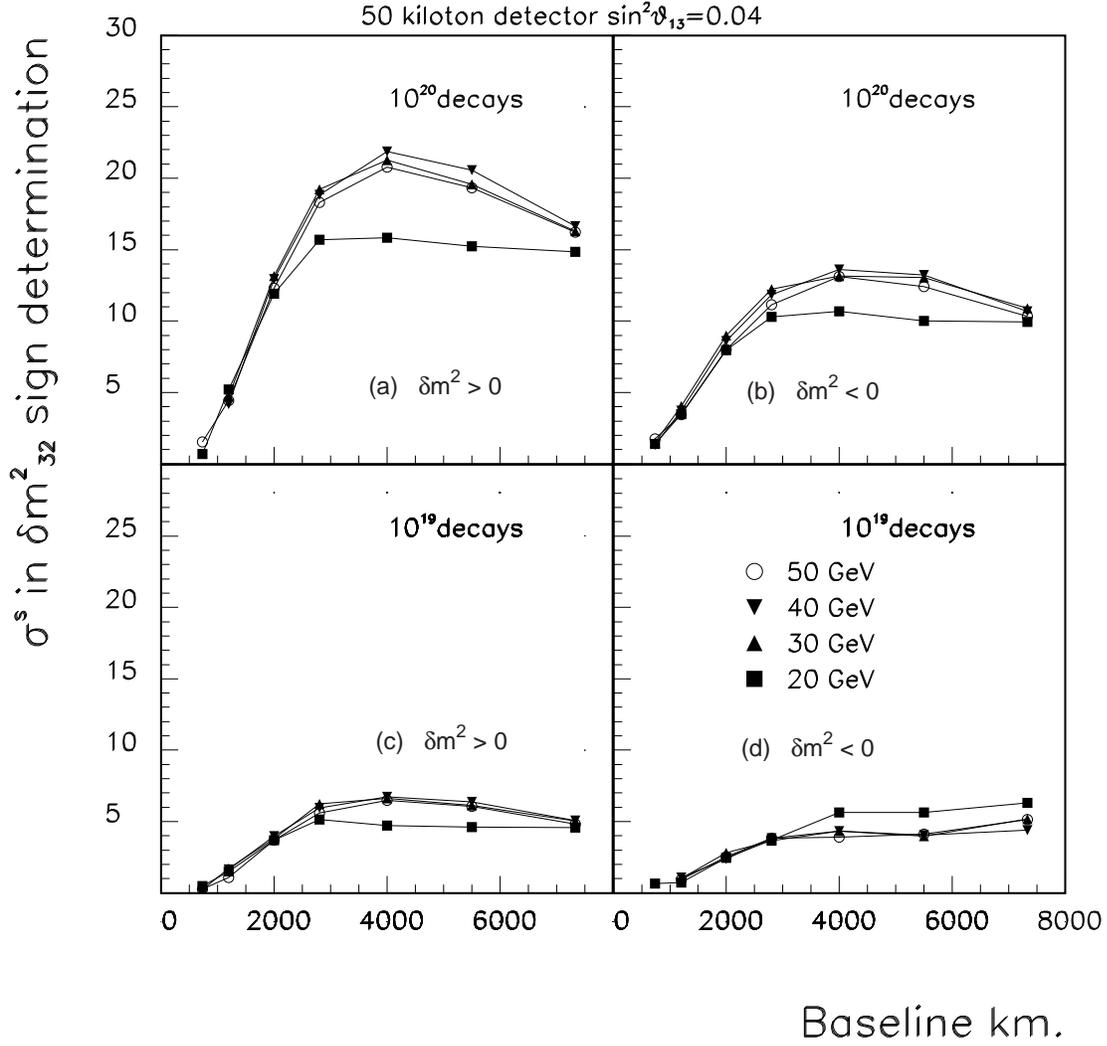}
\caption[]{The statistical significance (number of standard deviations) 
with which the
sign of $\delta m_{32}^2$ can be determined versus baseline length for
various muon storage ring energies. The results are shown for 
a 50~kiloton detector, and (a)~10$^{20}$
$\mu^+$ and $\mu^-$ decays and positive values of $\delta m_{32}^2$;
(b)~10$^{20}$ $\mu^+$ and $\mu^-$ decays and
negative values of $\delta m_{32}^2$; (c)~10$^{19}$ $\mu^+$ and 
$\mu^-$ decays and positive values of $\delta
m_{32}^2$; (d)~10$^{19}$ $\mu^+$ and $\mu^-$
decays and negative values of $\delta m_{32}^2$.
\label{fig:sigmas}}
\end{figure}

Figure \ref{fig:sigmas} shows the difference in negative
log-likelihood between a
correct and wrong-sign mass hypothesis expressed as a number of
equivalent Gaussian standard deviations versus baseline length for
muon storage ring energies of 20, 30, 40 and 50~GeV. The values of the
oscillation parameters are for the LMA scenario in
Eq.~\ref{eq:lam}. Figure \ref{fig:sigmas}(a) is for 10$^{20}$ decays
for each sign of stored energy and a 50 kiloton detector and positive
$\delta m^2_{32}$ (b) for negative $\delta m^2_{32}$ for various
values of stored muon energy. Figures \ref{fig:sigmas} (c) and (d)
show the corresponding curves for 10$^{19}$ decays and a 50 kiloton
detector. An entry-level machine would permit one to perform a
5$\sigma$ differentiation of the sign of $\delta m^2_{32}$ at a
baseline length of $\sim$2800~km.

\begin{figure}
\epsfxsize=6in\epsffile{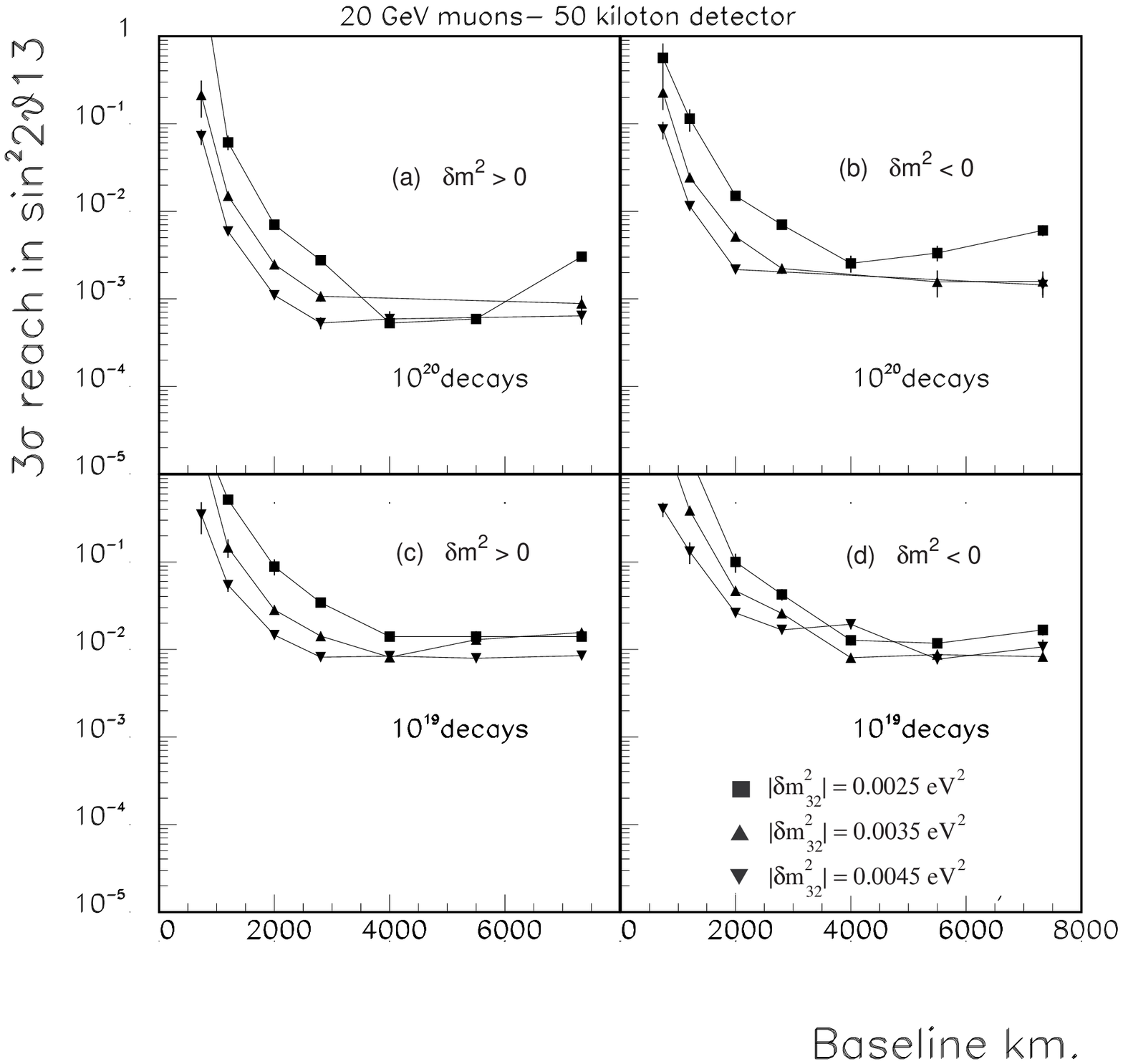}
\caption[]{The minimum value of $\sin^22\theta_{13}$ for which the
sign of $\delta m_{32}^2$ can be determined to at least 3 standard deviations
versus baseline length for various values of $\delta m_{32}^2$.
The results are shown for a 50~kiloton detector, and (a)~10$^{20}$ 
$\mu^+$ and $\mu^-$ decays and positive
values of $\delta m_{32}^2$; (b)~10$^{20}$
$\mu^+$ and $\mu^-$ decays and negative values of $\delta m_{32}^2$;
(c)~10$^{19}$ $\mu^+$ and $\mu^-$ decays and
positive values of $\delta m_{32}^2$; (d)~10$^{19}$ 
$\mu^+$ and $\mu^-$ decays and negative values of $\delta m_{32}^2$.
\label{fig:sensitivity}}
\end{figure}

Figure \ref{fig:sensitivity} shows the 3$\sigma$ reach of
$\sin^22\theta_{13}$ versus baseline length for a 20 GeV muon storage
ring (a)~with $10^{20}$ decays, a 50 kiloton detector and positive
$\delta m^2_{32}$, (b)~for negative $\delta m^2_{32}$ for various
values of $|\delta m^2_{32}|$. Figures
\ref{fig:sensitivity} (c) and (d) show the corresponding curves for
10$^{19}$ decays and a 50 kiloton detector. The error bars show the
uncertainties due to statistical fluctuations in our determination of the
3$\sigma$ reach. Our results agree with similar calculations in
Ref.~\cite{freund}. It can be seen that an entry-level machine is
capable of determining the sign of $\delta m^2_{32}$ provided that
$\sin^22\theta_{13}$ is greater than 0.01 for $\delta m^2_{32}$ in the
range 0.0025--0.0045 eV$^2$.

To conclude, we have shown that the neutrino factory provides a
powerful tool to explore the structure of neutrino masses. Even an
``entry level'' machine with $10^{19}$ decays of 20 GeV stored muons
and a 50 kiloton detector at a baseline exceeding $\sim$2800 km is capable
of differentiating the sign of $\delta m^2_{32}$ for the LMA
scenario. The $3\sigma$ reach in $\sin^2 2\theta_{13}$ in the sign
determination of $\delta m^2_{32}$ for such a machine exceeds the
present bound on $\sin^2 2\theta_{13}$ by about an order of magnitude.

\section*{Acknowledgments}

This research was supported in part by the U.S.~Department of Energy
under Grant No.~DE-FG02-95ER40896 and in part by the University of
Wisconsin Research Committee with funds granted by the Wisconsin
Alumni Research Foundation.

\newpage

%1

%2

%3

%4


\begin{thebibliography}{99}
\bibitem{decays}In principle, one can optimize further by running for
different periods of time for the different stored charges, to take
into account the difference in cross sections for neutrinos and
anti-neutrinos in matter, but we have not performed that optimization
in this study.
\end{thebibliography}
\end{document}